%% file: submit.tex
\documentclass{article}         
\usepackage{graphicx}
\usepackage{subfigure}
\usepackage{color}
\usepackage{cite}
\usepackage{clrscode}
\usepackage{amsmath}
\usepackage{fullpage}

\let\epsilon=\varepsilon

\newcommand{\bi}{\begin{itemize}}
\newcommand{\ei}{\end  {itemize}}
\newcommand{\bt}{\begin{tabbing}}
\newcommand{\et}{\end  {tabbing}}
\newcommand{\be}{\begin{enumerate}}
\newcommand{\ee}{\end  {enumerate}}

\newtheorem{theorem}{Theorem}

\begin{document}

\title{Optimal Free-Space Management and Routing-Conscious Dynamic Placement for Reconfigurable Devices\thanks{A short abstract summarizing the results of this paper appeared in 
the {\em Proceedings of the 14th International Conference on Field-Programmable Logic and Application (FPL '04)}\cite{abftv-orcdprd-04}.} }

\author{
Ali~Ahmadinia\thanks{Department of Computer Science 12, University of Erlangen-Nuremberg, Germany.  E-mail: \{ahmadinia, bobda, teich\}@cs.fau.de}
\and Christophe~Bobda\footnotemark[2]
\and S\'andor~P.~Fekete\thanks{Department of Mathematical Optimization, Braunschweig University of Technology, Germany. E-mail:\{s.fekete,j.van-der-veen\}@tu-bs.de}
\and  J\"urgen~Teich\footnotemark[2]
\and Jan~C.~van~der~Veen\footnotemark[3]
}

\date{}

\maketitle


\begin{abstract}
  We describe algorithmic results on two crucial aspects of allocating
  resources on computational hardware devices with partial
  reconfigurability. By using methods from the field of computational
  geometry, we derive a method that allows correct maintainance of
  free and occupied space of a set of $n$ rectangular modules in time
  $O(n\log n)$; previous approaches needed a time of $O(n^2)$ for
  correct results and $O(n)$ for heuristic results. We also show a
  matching lower bound of $\Omega(n\log n)$, so our approach is optimal.
  We also show that finding an optimal feasible
  communication-conscious placement (which minimizes the total
  weighted Manhattan distance between the new module and existing
  demand points) can be computed with $\Theta(n\log n)$. Both
  resulting algorithms are practically easy to implement and show
  convincing experimental behavior.
\end{abstract}

{\bf ACM Classification:} C.3.e: Reconfigurable Hardware; \\
\hspace*{5cm}F.2.2.c: Geometrical problems and computations

{\bf Keywords:} Reconfigurable hardware, field-programmable gate array (FPGA), 
module placement, free space manager, routing-conscious placement, 
geometric optimization, line sweep technique, optimal running time, 
lower bounds.

\input{Introduction}
\input{FreeSpaceManager}

\input{Placer}

\input{ExperimentalResults}

\input{Conclusion}
\bibliographystyle{IEEEtran}
\newpage
\bibliography{rc,geom}

\end{document}

%% file: Introduction.tex
\section{Introduction}
\label{sec:intro}

%
%
%

\paragraph{Reconfigurable Computing.}
{O}{ne} of the cutting-edge aspects of modern reconfigurable computing
is the possibility of {\em partial} reconfiguration of a device: A new
module can be placed on a reconfigurable chip whithout interfering the
computation of other modules. Clearly, this approach has advantages
over a full reconfiguration of the whole chip. However, there is still
a tremendous need for scientific progress: The technical possibilities
for partial reconfiguration have been somewhat restricted, and
manufacturers have been slow in providing possibilities, tools, and
documentation. As a consequence, there has only been a limited amount
of previous research on this topic.

New reconfigurable devices such as FPGAs offer increasing
levels of partial reconfigurability, and chip sizes continue to grow. 
At the same time, static programming methodologies
show an increasing use of pre-implementation by means of relocatable module 
libraries with bounding-box restrictions.

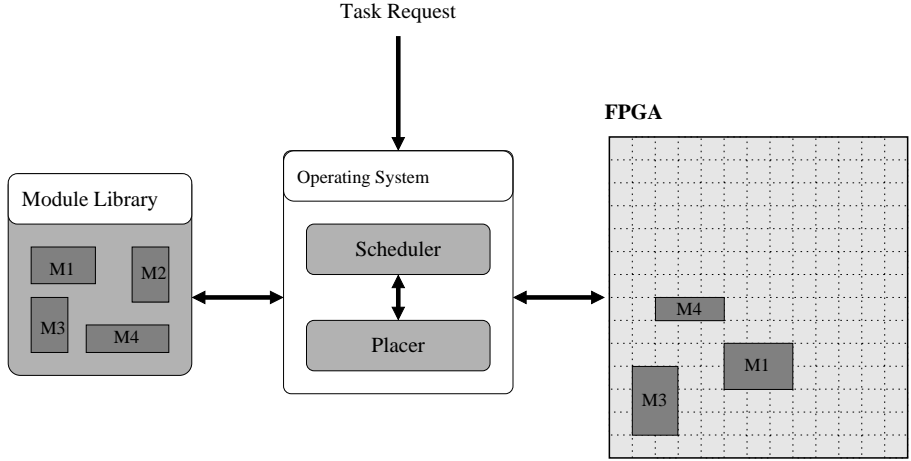
\begin{figure}[htb]
  \centering
  \input{betriebssystem.pstex_t}
  \caption{\em An operating system for reconfigurable computers.}
  \label{fi:opsys}
\end{figure}

These developments place an ever-growing demand on
the run-time management of resource allocation.
As these tasks become more and more complex, one needs
support in the form of operating systems \cite{wk-frosrc-01}
for managing both software and hardware processes (see Figure~\ref{fi:opsys}.)

Runtime space allocation, also known as temporal placement,
is a central part in partially reconfigurable computing systems.
In this paper, we present methods to solve two crucial issues for
devices that allow for partial reconfiguration:

\begin{enumerate}
\item Given a set of $n$ rectangular modules that have been placed
on a chip, identify all feasible positions for a new module.
\item Given a set of $n$ rectangular modules that have been placed on
  a chip, a new module, and demands for connecting it to existing
  sites, find a feasible position for the module that minimizes the
  total weighted distance to the given sites.
\end{enumerate}


{\bf Related Work.}  The first of the above issues is the task of
maintaining free space. Bazargan et al. \cite{bazargan00fast} describe
how to achieve this by maintaining the set of all maximal free
rectangles; as this set can have size $\Omega(n^2)$, the complexity is
quadratic. Alternatively, they propose partitioning free space into
only $O(n)$ free rectangles; the price for this improved complexity is
the fact that no feasible placement may be found, even though one
exists.  Walder et al.~\cite{WaStPl03} have suggested ways to reduce
this deficiency and did report on experimental improvement, but their
$O(n)$ procedure is still a heuristic approach that may fail in some
scenarios.
Thus, there remains a gap between $O(n^2)$ methods that report
an accurate answer, and $O(n)$ heuristics that may fail
in some scenarios. 
Ahmadinia et al.~\cite{ABT2004} suggested maintaining occupied
space instead of free space, but (depending on the computational
model) their approach is still quadratic.
Other current work on free-space management was presented by
Tabero et al.~\cite{tsmm-vla2d-03,tsmm-lfhtp-04}, who provide an $O(n^2)$ 
approach based on keeping track of possible corner positions.

The above difficulty may have contributed to the fact that
routing-con\-sci\-ous placement has received hardly any attention at all:
Clearly, {\em optimal} placement of a new module has to go beyond
feasible placement. In the context of configurable computing, this
second aspect has only been treated very recently, in work by
Ahmadinia et al.~\cite{ABT2004}, who suggest a heuristic to find a
feasible placement for a new module that has small total weighted
Euclidean distance to a given set of demand points. In the area of
discrete algorithms, two papers study a somewhat related problem: Karp
et al.~\cite{KarpMcWo75} consider the problem of arranging a set of
records in a 2-dimensional array, such that the total weighted
distance is minimized.  In this context, all records have the same
size (unit cells), and no previous records have been placed; on the
other hand, all records have to be placed at once, which is different
from our scenario. It should be noted that for the case of Manhattan
distances, the resulting shape for large numbers of records can only
be described by a differential equation, indicating surprising
computational difficulties.  In more recent work, Bender et
al.~\cite{bbd+-capas-04} consider the problem of allocating $k$
processors in a grid supercomputer in the presence of occupied cells;
this is a generalization of \cite{KarpMcWo75}. They also present
empirical evidence that indeed the Manhattan distance between
processors should be minimized for optimizing communication cost, and
thus runtime of the resulting jobs.  In this context, see also
\cite{krumke97,mache96,mache97b}.

\vspace*{2ex} {\bf Our Results.}  
We resolve both of the above issues:
\vspace*{-1ex}
\begin{itemize}
\item We give a $O(n\log n)$ method to provide a free-space
  manager (FSM).  This approach uses a plane-sweep approach from
  computational geometry.
\item We give a matching lower bound of $\Omega(n\log n)$ for locating
a maximal free rectangle between a set of $n$ modules, showing that our
method has optimal complexity.
\item We show that our FSM can be extended to find a feasible position
  that minimizes total weighted Manhattan distance to existing sites.
  The resulting algorithm still has an optimal run time of
  $\Theta(n\log n)$.
\item We describe implementation details to illustrate that our method
  is fast and easy.
\item We provide experimental data to demonstrate the practical usefulness
of our results.
\end{itemize}

The rest of this paper is organized as follows.  In
Section~\ref{sec:fsm}, we present our optimal FSM.
Section~\ref{sec:opt} describes how to perform optimal
routing-conscious placement. Section~\ref{sec:exp} shows
implementation details and experimental data. The final
Section~\ref{sec:Conclusion} discusses possible implications and
extensions of our work.


%% file: betriebssystem.pstex_t
\begin{picture}(0,0)%
\includegraphics{betriebssystem.pstex}%
\end{picture}%
\setlength{\unitlength}{2528sp}%
\begingroup\makeatletter\ifx\SetFigFont\undefined%
\gdef\SetFigFont#1#2#3#4#5{%
  \reset@font\fontsize{#1}{#2pt}%
  \fontfamily{#3}\fontseries{#4}\fontshape{#5}%
  \selectfont}%
\fi\endgroup%
\begin{picture}(8854,4498)(3139,-7598)
\end{picture}

%% file: FreeSpaceManager.tex
\section{The Free-Space Manager}
\label{sec:fsm}

In this section we present our approach to free-space management. Our
FSM is based on the observation that the occupied space consists of
very simple geometric objects, namely $n$ placed rectangular modules.
Put simply our free-space manager is a modification of the well-known
algorithm \textsc{ContourOfUnionOfRectangles} (CUR)
\cite{g-ocaio-84,lp-fcuio-80,ps-cgi-85}, for finding the contour of a
union of axis-parallel rectangles.  As the number of contour segments
is linear in $n$, we achieve a running time of $O(n \log n)$.  Note
that we do {\em not} require the contour to be connected, i.e., our
approach works even if there are holes in the arrangement.

\subsection{Free-Space Manager Basics}

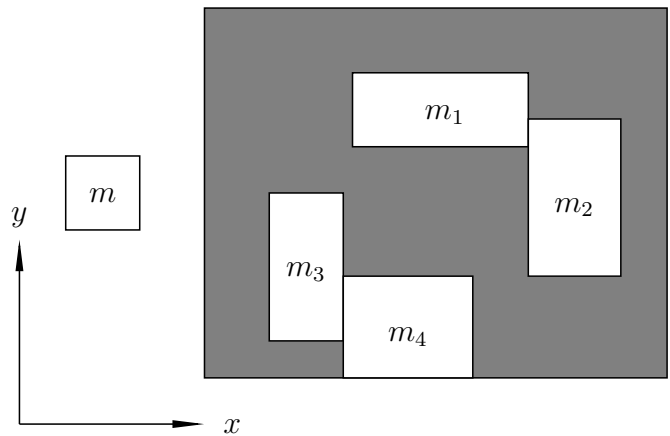
\begin{figure}[htb]
  \centering
  \input{PlaceRectangle.pstex_t}
  \caption{\em A set of existing modules (white) on a rectangular
chip and an additional module that has to be placed on the remaining
free chip area (dark).}
  \label{fig:BeforeBlowUp}
\end{figure}

We consider an FPGA or other reconfigurable device and denote its
width by $W$ and its height by $H$.  Assumsing a coordinate origin in
the lower left corner of the chip, we can describe the corresponding
input by the quadruple $F = (0, 0, W, H)$.  On the device, a set $M =
\{(x_i, y_i, w_i, h_i) : i \in \{1, 2, \ldots, n\}\}$ of modules $m_i$
with widths $w_i$ and heights $h_i$ has been placed, with lower left
corners at positions $(x_i, y_i)$.  The task for the free-space
manager is to identify regions where a new module $m$ with width $w_m$
and height $h_m$ can be placed.

\begin{figure}[htb]
  \centering
  \input{PlacePoint.pstex_t}
  \caption{\em Expanding existing modules and shrinking chip area and the new 
    module reduces free-space management to managing free-space for a
    single point.}
  \label{fig:AfterBlowUp}
\end{figure}
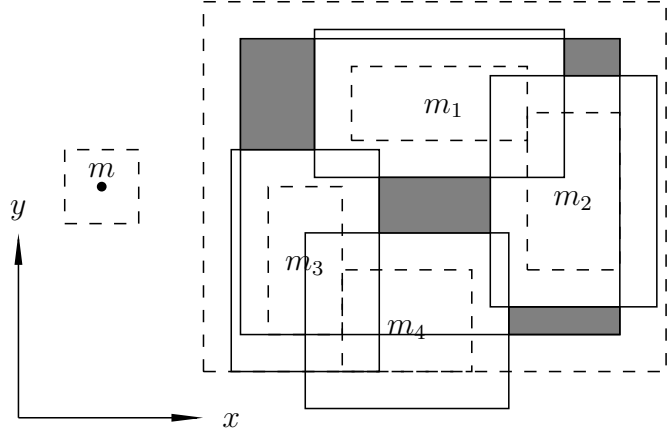

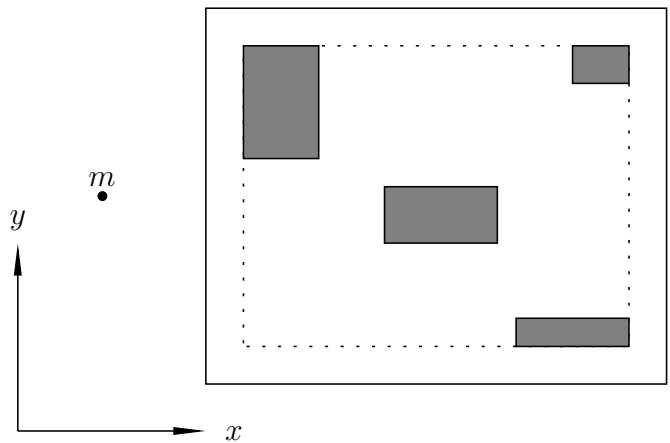
\begin{figure}[htb]
  \centering 
  \input{PlacePointUnion.pstex_t}
  \caption{\em Resulting free space (dark).}
  \label{fig:ResultingFreeSpace}
\end{figure}

As mentioned above, prior free-space managers maintained lists of
unoccupied rectangles. Because the number of maximal empty rectangles
is quadratic in the number of modules this clearly leads to quadratic
running times.  In \cite{ABT2004}, Ahmadinia et al.\ described how the
management of free space can be simplified to finding a placement for
a single point (see Figure~\ref{fig:AfterBlowUp}) by transforming the
problem data as follows: Shrink the area of the chip and
simultaneously blow up the existing modules by half the width and half
the height of the new module. This way the chip area is given by
\begin{align*}
  F' &= (\frac{w_m}{2}, \frac{h_m}{2}, W - \frac{w_m}{2}, H -
  \frac{h_m}{2})\\
  \intertext{and the set of placed modules $M$ clipped by $F'$ becomes}\\
  M' &= \{(x'_i, y'_i, w'_i, h'_i) : i \in \{1, 2, \ldots, n\}\}\\
  \intertext{with}\\
  x'_i &= \max\{x_i - \frac{w_m}{2}, \frac{w_m}{2}\},\\
  y'_i &= \max\{y_i - \frac{h_m}{2}, \frac{h_m}{2}\},\\
  w'_i &= \min\{w_i + w_m, W - w_m\},\\
  h'_i &= \min\{h_i + h_m, H - h_m\}.
\end{align*}

The new module $m$ reduces to a point $m'$ and the original problem of
finding free space for a rectangle reduces to finding free space for
that point.

For the module to be placed shown in Figure~\ref{fig:AfterBlowUp} the
result of the transformation is shown in
Figure~\ref{fig:ResultingFreeSpace}. All points within and on the
border of the shaded areas are feasible placement locations.

\subsection{Representation of Free Space}

Among all feasible placements, all points on the contour of the
free space are feasible. As we will see in the next
Section~\ref{sec:opt}, these positions are of particular interest when
trying to preserve a good structure of free space, and minimizing
total communication distance.

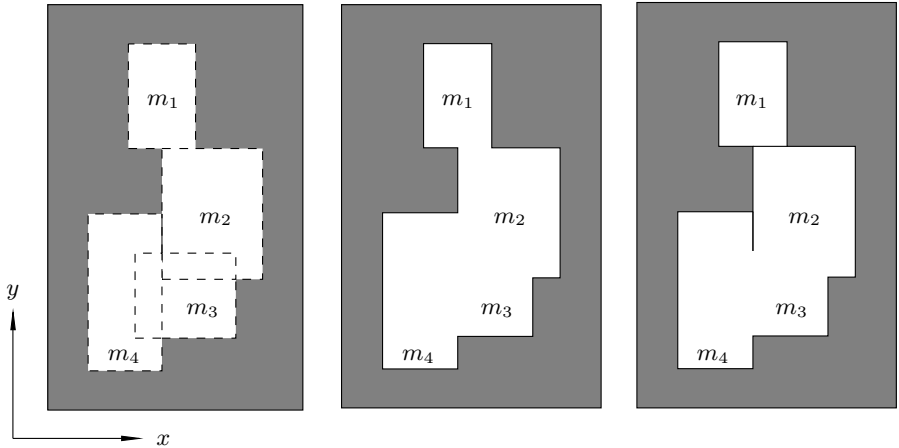
\begin{figure}[htb]
  \centering
  \input{ModifiedContour.pstex_t}
  \caption{\em (Left) A set of rectangles, representing expanded modules (dashed, light).
    (Center) The contour as returned by CUR, and the 
    resulting feasible space for placing the center point of the new module 
    (dark, with boundary).  (Right) Correct feasible
    positions for placing the center point of the new module, 
    as returned by our FSM (dark, with boundary.)}
  \label{fig:ModifiedContour}
\end{figure}

In general finding the contour of a set of $n$ axis-aligned rectangles
can be done in $O(n \log n + s)$ by using the CUR algorithm as
described in \cite{g-ocaio-84,lp-fcuio-80,ps-cgi-85}. Here $s$ is the
complexity of the resulting contour. Our algorithm is not simply an
implementation of CUR. There are a few subtlelties that have to be
considered. All differences stem from the above mentioned fact, that
the points on the contour are feasible placements. As a consequence
our algorithm has to find free space of height and width 0 (see Figure
\ref{fig:ModifiedContour}). In the following we will describe CUR and
our modifications to it.

The building blocks of CUR are an algorithmic technique from
computational geometry called \emph{plane sweep} and a data structure
called \emph{segment tree}.  For an in-depth introduction to both see
\cite{bkos-cgaa-00}.

A {\em plane-sweep algorithm} is an algorithm that scans the plane and
a set of objects in it: Move an axis-parallel line in an orthogonal
direction across the plane and keep track of the structure of the
intersection with the set of objects.  The key observation is to
notice that updates to this structure only occur at a discrete set of
critical positions called {\em events}.  By pre-sorting these events
(in time $\Theta(n\log n)$), only the updates have to be performed,
which can be done efficiently for all events by using an appropriate
data structure.  For our purposes, such a data structure is a {\em
  segment tree}: This is a balanced binary tree for dynamically
storing a set of $n$ intervals.  The number of endpoints of these
intervals must be known at construction time. Because it is bounded by
$2n$ the segment tree can be constructed in $O(n)$.  Insertion and
deletion of intervals can be done in $O(\log n)$.  Segment trees as
used in \cite{b-mbstu-75,ps-cgi-85} have been introduced in
\cite{b-skrp-77}. See \cite{bw-owcar-80} for more details.

One has to be careful when constructing the segment tree. To find free
space of height and width 0 we have to make sure that two modules
starting or ending on the same coordinate are separated by an
elementary interval in the segment tree. This can be done by
disturbing the top left corner of each module by a sufficiently small
$\epsilon > 0$.

The crucial part of our algorithm are two plane sweeps: one horizontal
sweep that discovers all the vertical contour segments and one
vertical sweep that finds all horizontal segments. As the
horizontal and the vertical sweep differ only in the initialization, we
only describe the horizontal sweep.

For the horizontal sweep we construct a list $L$ of $2n$ quadruples,
denoted by $(p_i, t_i, b_i, e_i)$: for each of the modules in $M'$, we add two
elements to $L$ --- one for the left side $(x'_i, Open, y'_i, y'_i +
h'_i)$ and one for the right side $(x'_i + w'_i, Close, y'_i, y'_i +
h'_i)$. This list is sorted lexicographically and we assume that
$Close < Open$. 

In the sweep we process all $2n$ elements of $L$. In case of an $Open$
event the corresponding contour points are retrieved from the
segment tree and the segment $[y'_j, y'_j + h'_j]$ is added to
the tree. For a $Close$ event the segment $[y'_i, y'_i + h'_i]$ is
removed from the tree and the corresponding contour points 
are retrieved. 


In the CUR algorithm we would construct the horizontal contour
segments from the vertical segments. In our setting we might not find
all free space of height $0$. So we need to do another vertical plane
sweep to discover all horizontal segments.



\subsection{Combinatorial Complexity of the Free-Space Contour}

In this section we will show that the combinatorial complexity of the
contour of the free space is linear in the number of modules. We
thereby show that the complexity of our algorithm is $O(n \log n)$.

\begin{figure}[htb]
 \begin{center}
  \leavevmode
  \includegraphics[width=.5\columnwidth]{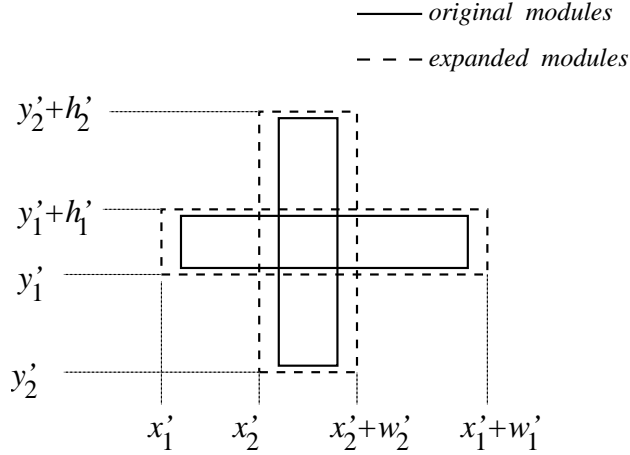}
  \caption{\em An impossible, crossing position for expanded modules.}
  \label{fig:Cross}
  \vspace*{-3mm}
 \end{center}
\end{figure}

In general, the contour of a union of $n$ rectangles may consist of
$\Omega(n^2)$ line segments, e.g., when considering two sets
of pairwise overlapping $n/2$ axis-parallel strips, where the intersections
form a grid pattern. This is not the case for the sets of rectangles
arising as expanded modules; in fact, we prove that an arrangement of
rectangles as in Figure~\ref{fig:Cross} is impossible. As this is the only
arrangement of two rectangles for which the number of edges
forming the contour exceeds eight, this can be used
as a stepping stone for an inductive proof that 
the contour never consists of more than $4n$ line segments.

\begin{theorem} \label{th:cross}
The expanded regions $m'_i, m'_j \in M'$ for
two existing, disjoint modules $m_i, m_j \in M$ cannot cross.
\end{theorem}
\begin{proof}
  Every expanded module $m'_i$ is a rectangle, described by four bounding
  coordinates:
  $x'_i$, the position of its left edge;
  $x'_i+w'_i$, the position of its right edge;
  $y'_i$, the position of its lower edge;
  $y'_i+h'_i$, the position of its upper edge.
  Now assume there are two expanded modules that cross,
  say, $m_1$ and $m_2$, and consider without loss of generality
  that $x'_1<x'_2$.
  Then crossing means that $x'_2+w'_2<x'_1+w'_2$, 
  $y'_2<y'_1$, and $y'_1+h'_1<y'_2+h'_2$.
  But as all expanded modules arise from the original modules by moving
  their edges by the same amount for each coordinate direction, 
  the same relative order must be valid for the original edges. 
  This implies that the original modules
  cross, contradicting their disjointness.
\end{proof}
\begin{theorem} \label{th:LinearNumberOfSegments}
  \proc{FindContourSegments} finds $O(n)$ contour segments.
\end{theorem}
\begin{proof}
  We will argue that the number of segments of the contour is at most
  the number of segments of all modules.
  
  Consider the frontier $O_i = \{[b_1, e_1], \ldots, [b_o, e_o]\}$ of
  all open segments just before encountering event $L_i = (p_i, t_i,
  b_i, e_i)$.
  
  Let us consider an $Open$ event $L_i$. Processing this event yields
  at most $|O_i|$ new contour segments. This is the case if and only
  if all elements $[b_j, e_j] \in O_i$ are completely contained in
  $[b_i, e_i]$ and do not overlap.  But because no two modules can
  cross, all $Close$ events $L_{j'}$ must occur before the
  $Close$ event $L_{i'}$.  Consequently the closing segments $[b_{j'},
  e_{j'}]$ do not contribute to the contour and in total the number of
  segments added is at most one.
  
  Now let us consider a $Close$ event $L_i$. Arguing as above, we
  conclude that if $[b_i, e_i]$ is totally contained in an element
  $[b_j, e_j] \in O_i$, the total number of segments added to the
  contour is at most one.
  
  In all other events at most one segment is added to the contour. As
  there are not more than $2n$ events in each cordinate direction, 
  the number of contour segments parallel to each axis is at most $2n$. 
  Thus, the number of contour segments is at most $4n$, i.e., linear in $n$. 
\end{proof}

\subsection{Computational Complexity}

\begin{theorem}
The complexity of \proc{FindContourSegments} is $O(n\log n)$.
\end{theorem}
\begin{proof}
  The algorithm CUR has a running time of $O(n \log n + s)$, where $s$
  is the number of segments of the contour. As we have shown in
  Theorem \ref{th:LinearNumberOfSegments}, $s = O(n)$. Our
  modifications to CUR do not increase the running time by more than a
  constant factor. Thus the running time of \proc{FindContourSegments}
  is $O(n \log n)$.
\end{proof}

\subsection{Lower Bound}
Assuming a standard computational model, we can show that our FSM has
optimal running time, by providing a matching lower bound:

\begin{theorem} \label{th:lower}
  In the algebraic tree model of computation, there is a lower bound
  of $\Omega(n\log n)$ on the complexity of deciding the maximum size
  of a free rectangle between $n$ existing rectangles.
\end{theorem}
\begin{proof}
  The claim is already true in one dimension, for $n$ existing unit
  intervals, with positions described by $n$ integers $a_1, \ldots,
  a_n$. Determining a maximum free interval is precisely the problem
  {\sc Maximum Gap}.
  
  The {\sc Maximum Gap} problem is the problem of determining the
  maximum gap between two consecutive numbers of a set $A = \{a_1, a_2,
    \ldots, a_n\}$ of $n$ numbers. Two elements of $A$ are called
  consecutive if they appear consecutively in the sequence obtained by
  sorting $A$. The running time of {\sc Maximum Gap} is bounded from
  below by $\Omega(n\log n)$, as described in Chapter 6 of
  \cite{ps-cgi-85}.
\end{proof}


%% file: PlaceRectangle.pstex_t
\begin{picture}(0,0)%
\includegraphics{PlaceRectangle.pstex}%
\end{picture}%
\setlength{\unitlength}{5097sp}%
\begingroup\makeatletter\ifx\SetFigFont\undefined%
\gdef\SetFigFont#1#2#3#4#5{%
  \reset@font\fontsize{#1}{#2pt}%
  \fontfamily{#3}\fontseries{#4}\fontshape{#5}%
  \selectfont}%
\fi\endgroup%
\begin{picture}(3189,2122)(-26,-1271)
\put(406,-106){\makebox(0,0)[b]{\smash{\SetFigFont{12}{14.4}{\rmdefault}{\mddefault}{\updefault}{\color[rgb]{0,0,0}$m$}%
}}}
\put(1036,-1231){\makebox(0,0)[b]{\smash{\SetFigFont{12}{14.4}{\rmdefault}{\mddefault}{\updefault}{\color[rgb]{0,0,0}$x$}%
}}}
\put(1396,-466){\makebox(0,0)[b]{\smash{\SetFigFont{12}{14.4}{\rmdefault}{\mddefault}{\updefault}{\color[rgb]{0,0,0}$m_3$}%
}}}
\put(1891,-781){\makebox(0,0)[b]{\smash{\SetFigFont{12}{14.4}{\rmdefault}{\mddefault}{\updefault}{\color[rgb]{0,0,0}$m_4$}%
}}}
\put(2701,-151){\makebox(0,0)[b]{\smash{\SetFigFont{12}{14.4}{\rmdefault}{\mddefault}{\updefault}{\color[rgb]{0,0,0}$m_2$}%
}}}
\put(2071,299){\makebox(0,0)[b]{\smash{\SetFigFont{12}{14.4}{\rmdefault}{\mddefault}{\updefault}{\color[rgb]{0,0,0}$m_1$}%
}}}
\put(  1,-196){\makebox(0,0)[b]{\smash{\SetFigFont{12}{14.4}{\rmdefault}{\mddefault}{\updefault}{\color[rgb]{0,0,0}$y$}%
}}}
\end{picture}

%% file: PlacePoint.pstex_t
\begin{picture}(0,0)%
\includegraphics{PlacePoint.pstex}%
\end{picture}%
\setlength{\unitlength}{5097sp}%
\begingroup\makeatletter\ifx\SetFigFont\undefined%
\gdef\SetFigFont#1#2#3#4#5{%
  \reset@font\fontsize{#1}{#2pt}%
  \fontfamily{#3}\fontseries{#4}\fontshape{#5}%
  \selectfont}%
\fi\endgroup%
\begin{picture}(3189,2122)(-26,-1271)
\put(406,-16){\makebox(0,0)[b]{\smash{\SetFigFont{12}{14.4}{\rmdefault}{\mddefault}{\updefault}{\color[rgb]{0,0,0}$m$}%
}}}
\put(1036,-1231){\makebox(0,0)[b]{\smash{\SetFigFont{12}{14.4}{\rmdefault}{\mddefault}{\updefault}{\color[rgb]{0,0,0}$x$}%
}}}
\put(1396,-466){\makebox(0,0)[b]{\smash{\SetFigFont{12}{14.4}{\rmdefault}{\mddefault}{\updefault}{\color[rgb]{0,0,0}$m_3$}%
}}}
\put(1891,-781){\makebox(0,0)[b]{\smash{\SetFigFont{12}{14.4}{\rmdefault}{\mddefault}{\updefault}{\color[rgb]{0,0,0}$m_4$}%
}}}
\put(2701,-151){\makebox(0,0)[b]{\smash{\SetFigFont{12}{14.4}{\rmdefault}{\mddefault}{\updefault}{\color[rgb]{0,0,0}$m_2$}%
}}}
\put(2071,299){\makebox(0,0)[b]{\smash{\SetFigFont{12}{14.4}{\rmdefault}{\mddefault}{\updefault}{\color[rgb]{0,0,0}$m_1$}%
}}}
\put(  1,-196){\makebox(0,0)[b]{\smash{\SetFigFont{12}{14.4}{\rmdefault}{\mddefault}{\updefault}{\color[rgb]{0,0,0}$y$}%
}}}
\end{picture}

%% file: PlacePointUnion.pstex_t
\begin{picture}(0,0)%
\includegraphics{PlacePointUnion.pstex}%
\end{picture}%
\setlength{\unitlength}{5180sp}%
\begingroup\makeatletter\ifx\SetFigFont\undefined%
\gdef\SetFigFont#1#2#3#4#5{%
  \reset@font\fontsize{#1}{#2pt}%
  \fontfamily{#3}\fontseries{#4}\fontshape{#5}%
  \selectfont}%
\fi\endgroup%
\begin{picture}(3144,2122)(-26,-1271)
\put(406,-16){\makebox(0,0)[b]{\smash{\SetFigFont{12}{14.4}{\rmdefault}{\mddefault}{\updefault}{\color[rgb]{0,0,0}$m$}%
}}}
\put(1036,-1231){\makebox(0,0)[b]{\smash{\SetFigFont{12}{14.4}{\rmdefault}{\mddefault}{\updefault}{\color[rgb]{0,0,0}$x$}%
}}}
\put(  1,-196){\makebox(0,0)[b]{\smash{\SetFigFont{12}{14.4}{\rmdefault}{\mddefault}{\updefault}{\color[rgb]{0,0,0}$y$}%
}}}
\end{picture}

%% file: ModifiedContour.pstex_t
\begin{picture}(0,0)%
\includegraphics{ModifiedContour.pstex}%
\end{picture}%
\setlength{\unitlength}{3605sp}%
\begingroup\makeatletter\ifx\SetFigFont\undefined%
\gdef\SetFigFont#1#2#3#4#5{%
  \reset@font\fontsize{#1}{#2pt}%
  \fontfamily{#3}\fontseries{#4}\fontshape{#5}%
  \selectfont}%
\fi\endgroup%
\begin{picture}(6113,3097)(-250,-2261)
\put(811,119){\makebox(0,0)[b]{\smash{\SetFigFont{9}{10.8}{\rmdefault}{\mddefault}{\updefault}{\color[rgb]{0,0,0}$m_1$}%
}}}
\put(1171,-691){\makebox(0,0)[b]{\smash{\SetFigFont{9}{10.8}{\rmdefault}{\mddefault}{\updefault}{\color[rgb]{0,0,0}$m_2$}%
}}}
\put(541,-1636){\makebox(0,0)[b]{\smash{\SetFigFont{9}{10.8}{\rmdefault}{\mddefault}{\updefault}{\color[rgb]{0,0,0}$m_4$}%
}}}
\put(1081,-1321){\makebox(0,0)[b]{\smash{\SetFigFont{9}{10.8}{\rmdefault}{\mddefault}{\updefault}{\color[rgb]{0,0,0}$m_3$}%
}}}
\put(2836,119){\makebox(0,0)[b]{\smash{\SetFigFont{9}{10.8}{\rmdefault}{\mddefault}{\updefault}{\color[rgb]{0,0,0}$m_1$}%
}}}
\put(3196,-691){\makebox(0,0)[b]{\smash{\SetFigFont{9}{10.8}{\rmdefault}{\mddefault}{\updefault}{\color[rgb]{0,0,0}$m_2$}%
}}}
\put(2566,-1636){\makebox(0,0)[b]{\smash{\SetFigFont{9}{10.8}{\rmdefault}{\mddefault}{\updefault}{\color[rgb]{0,0,0}$m_4$}%
}}}
\put(3106,-1321){\makebox(0,0)[b]{\smash{\SetFigFont{9}{10.8}{\rmdefault}{\mddefault}{\updefault}{\color[rgb]{0,0,0}$m_3$}%
}}}
\put(4861,119){\makebox(0,0)[b]{\smash{\SetFigFont{9}{10.8}{\rmdefault}{\mddefault}{\updefault}{\color[rgb]{0,0,0}$m_1$}%
}}}
\put(5221,-691){\makebox(0,0)[b]{\smash{\SetFigFont{9}{10.8}{\rmdefault}{\mddefault}{\updefault}{\color[rgb]{0,0,0}$m_2$}%
}}}
\put(4591,-1636){\makebox(0,0)[b]{\smash{\SetFigFont{9}{10.8}{\rmdefault}{\mddefault}{\updefault}{\color[rgb]{0,0,0}$m_4$}%
}}}
\put(5131,-1321){\makebox(0,0)[b]{\smash{\SetFigFont{9}{10.8}{\rmdefault}{\mddefault}{\updefault}{\color[rgb]{0,0,0}$m_3$}%
}}}
\put(811,-2221){\makebox(0,0)[b]{\smash{\SetFigFont{9}{10.8}{\rmdefault}{\mddefault}{\updefault}{\color[rgb]{0,0,0}$x$}%
}}}
\put(-224,-1186){\makebox(0,0)[b]{\smash{\SetFigFont{9}{10.8}{\rmdefault}{\mddefault}{\updefault}{\color[rgb]{0,0,0}$y$}%
}}}
\end{picture}

%% file: Placer.tex
\section{Routing-Conscious Placement}
\label{sec:opt}

After describing how to find {\em all feasible} placements for a new
module, we turn to finding an {\em optimal} placement, such that the
weighted communication cost is minimized.  As described in the
introduction, an appropriate measure for this cost is the Manhattan
distance between modules, weighted by the relative amount of
communication.  This can still be achieved in time $\Theta(n\log n)$,
making use of local optimality properties, our FSM, and another
application of plane sweep techniques.

\subsection{Model}

Given $F'$ and $M'$ as in Section \ref{sec:fsm}, the objective of the
placer is to find a point in free space that minimizes communication
cost for the new module $m$.

\begin{figure}[htb]
  \centering
  \includegraphics[width=\columnwidth]{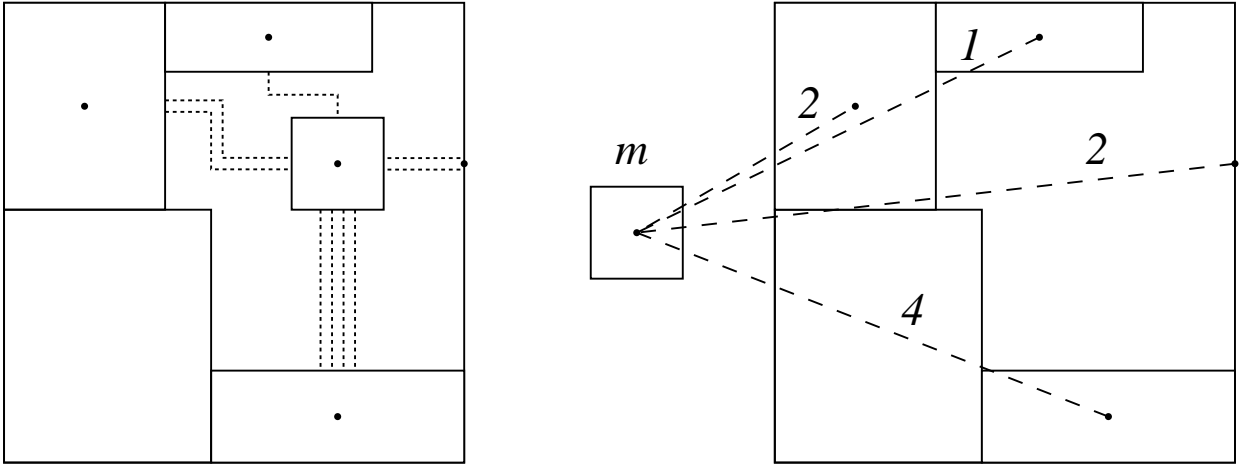}
  \caption{\em (Left) Physical chip with one module and its 
    connectivity to other modules drawn by dotted lines (Right) Dotted
    lines with bandwidth attribute $b_i^m$ to denote required
    connectivity of module $m$ to other modules $i$.}
  \label{fig:CommunicationModel}
\end{figure}

In our routing-conscious approach we let the communication cost of an
additional module $m$ depend on the distance to the centers of a
subset of existing modules, which is measured in the Manhattan metric.
We may also consider communication with the chip boundary as indicated
in Figure~\ref{fig:CommunicationModel}. So we have to consider the set
$C = \{(x_1, y_1), \ldots, (x_k, y_k)\}$ of demand points for
communication.  Clearly, we may assume that the number $k$ of
connections to be established with module $m$ is linear in $n$. A
second factor in communication cost is the width $b^m_i$ of the
communication path needed to create a routing unit between modules $i$
and $m$; this needs to be taken into account as a multiplicative
factor. Thus, we get the objective function
\begin{align*}
  &\min \{c(x'_m, y'_m) : (x'_m, y'_m) \in F' \setminus \bigcup_{m'_i \in M'} m'_i\} \\
  \intertext{with} &c(x, y) = \sum_{i=1}^{k} b^m_i\|(x_i, y_i) - (x,
  y)\|_1.  \intertext{Because we are dealing with the Manhattan
    metric, this can be reformulated to}
  &c(x, y) = c^x(x) + c^y(y) \\
  \intertext{with} &c^x(x)=\sum_{i=1}^{k}b^m_i|x_i-x|,
  \intertext{and} &c^y(y)=\sum_{i=1}^{k}b^m_i|y_i-y|.  \intertext{If
    we would allow $m'$ to be placed anywhere on the chip this can be
    reformulated to}
  &\min \{c^x(x'_m) : x'_m \in [{\frac{w_m}{2}}, W\!-\!{\frac{w_m}{2}}]\}+\\
  &\min\{c^y(y'_m) : y'_m \in [{\frac{h_m}{2}}, H\!-\!{\frac{h_m}{2}}]\}.\\
  \intertext{As a consequence we can consider two separate minimization
    problems, one for each coordinate. If we ignore feasibility, both
    minima are attained in the respective weighted medians, so they
    can be computed in linear time \cite{Blum:1972:LTB}; as we already
    sort the coordinates for performing plane sweep, this running time
    is not critical, so we may as well use a trivial method.  Note
    that only medians satisfy unconstrained local optimality, as the
    gradients for $c^x$ and $c^y$ are simply}
  &\nabla c^x(x)=\sum_{x_i<x}b^m_i-\sum_{x_i>x}b^m_i,\\
  \intertext{the sum of the required bandwidths to the left minus the
    sum of the required bandwidths to the right and} 
  &\nabla c^y(y)=\sum_{y_i<y}b^m_i-\sum_{y_i>y}b^m_i,
\end{align*}
the sum of the required buswidths to the bottom minus the sum of the
required buswidths to the top.  

\subsection{Local Optimality}

As we have seen in the previous subsection the median is the globally
optimal point if it is not in the occupied space. If it is in the
occupied space there are only two other types of points where the
global optimum could be located.

\begin{figure}[htb]
  \centering
  \includegraphics[width=.7\columnwidth]{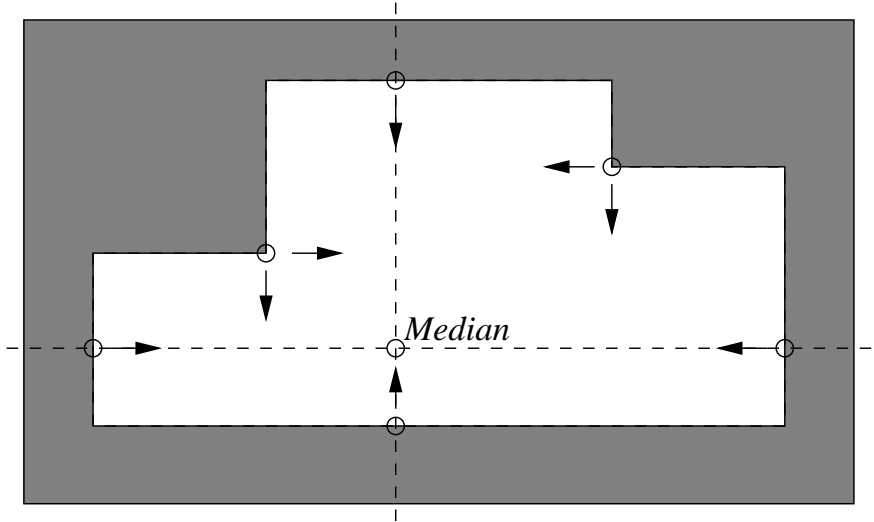}
  \caption{\em All potentially optimal points are marked by circles. As
    the median is located in the occupied space one of the other
    points must be optimal. The arrows show the directions in which
    the solution value would improve.}
  \label{fig:LocalOptimality}
\end{figure}

All points of one type can be found by intersecting the contour of the
occupied space with the median axes $l_x = \{(x_{med}, y) : y \in [0,
H]\}$ and $l_y = \{(x, y_{med}) : x \in [0, W]\}$. In these points one
of the gradients $\nabla c^x$ and $\nabla c^y$ vanishes. We cannot
move in the direction of a better solution because that way is blocked
by either a vertical or a horizontal segment of the contour.

The other type of points are some of the vertices of the contour.
These points are the intersections of horizontal and vertical segments
forming an interior angle of $\frac{\pi}{2}$ pointing in the direction of the
median. In these points neither of the gradients vanishes. Either
of the directions indicated by the gradient are blocked by contour
segments.

By simply inspecting all the local optima one finds the global
optimum. In the next subsection we describe how this can be done
efficiently.

\subsection{Algorithm for Global Optimality}

The FSM described in the previous section computes the contour of the
occupied space in $O(n \log n)$. A simple algorithm that finds the
optimal point to place a new module $m$ would compute the median and
check its feasibility; if the outcome is positive, we have found the
optimum.  Otherwise we need to check communication cost for all other
possible local minima, i.e., for every vertex of the contour and every
intersection point of the contour with one of the median axes. Let $L$
denote this set of points.  Computing commuication cost for a single
point takes $O(n)$, so evaluating all objective values in a
brute-force manner would take $O(n^2)$ time. However, by means of two
more plane sweeps, we can achieve a complexity of $O(n \log n)$.

For this purpose, we observed that communication cost for the $x$- and
$y$-coordinate of the contour segments can be computed separately,
then add the precomputed values for every point of $L$.  The crucial
step is to use the fact that we only need to compute the communication
cost for the leftmost $x$-coordinate and for the bottommost
$y$-coordinate; the other values can be obtained by doing appropriate
fast updates during the plane sweep.  Below we give details of this
step; for ease of presentation, we add the communcation points $C$ to
$L$ --- $L' = L \cup C$ and only describe updates for $x$-coordinates.

First we sort (in time $O(n\log n)$) $L'$ by increasing
$x$-coordinate.  Next we remove all points that are located on the
same $x$-coordinate in $O(n)$ time. With the convention that $b^m_i =
0$ if $(x_i, y_i) \in L \setminus C$ we compute the required bandwidth
to the left and to the right of each point $i$ by $l_i = 0$, $l_i =
l_{i-1} + b^m_i$, $r_{|L'|} = 0$, and $r_i = r_{i+1} + b^m_i$. These
values can be computed by a forward and a backward scan in $O(n)$
time. This yields the following recursive formula
\[c^x_{i + 1} = c_i + (l_i - r_{i+1})(x_{i+1}-x_i)\]
for computing communication cost for all $x$-coordinates 
in time linear in $n$.


As the lower bound from the previous section still applies,
we get the following:

\begin{theorem} \label{th:opt}
  A feasible position with minimum communication cost can be computed
  in time $\Theta(n\log n)$.
\end{theorem}


%% file: ExperimentalResults.tex
\vspace*{-8mm}
\section{Experimental Results}
\label{sec:exp}

\begin{table*}
\begin{center}
{\small
\begin{tabular}{|lr||r|r|r||r|r|r||r|r|r|}
\hline
& & \multicolumn{3}{l||}{Running Time (ms)} & \multicolumn{3}{l||}{Routing Cost} & \multicolumn{3}{l|}{Rejection Rate} \\
       &       & RCP & NAO & KFF &  RCP & NAO &   KFF &  RCP & NAO & KFF \\
\hline
\hline 
Uniform   & 5-10\%& 173 &  197 & 204 & 1403 & 3641 &  9522 &  0\% &  0\% & 0\% \\
\hline
Uniform   &10-15\%& 162 &  208 & 194 & 1747 & 5490 & 14311 &  0\% &  1\% & 0\% \\
\hline
Uniform   &15-20\%& 160 &  172 & 158 & 2044 & 7250 & 19791 &  2\% &  5\% & 1\% \\
\hline
Uniform   &20-25\%& 156 &  181 & 161 & 1987 & 7061 & 20159 & 10\% & 12\% & 9\% \\
\hline
\hline
Uniform   & 5-25\%& 168 &  224 & 215 & 1721 & 6741 & 21347 &  5\% &  8\% & 5\% \\
\hline
Increasing& 5-25\%& 196 &  252 & 243 & 1931 & 6914 & 21910 &  8\% & 14\% & 6\% \\
\hline
Decreasing&25-5\% & 175 &  232 & 228 &  611 & 2311 & 11712 &  0\% &  3\% & 4\% \\
\hline
\hline
Average   &       & 170 &  209 & 200 & 1635 & 5630 & 16965 &  3.6\% &  6.1\% & 3.6\% \\
\hline
\end{tabular}
}
\par\medskip
\caption{\em \label{bench}Experimental results for the different benchmark
  instances. Overall running time, average routing cost for each
  module, and rejection rate are shown for the different algorithms.
  RCP denotes the algorithm described in this paper, NAO refers to
  the algorithm as described in \cite{ABT2004} and KFF is the
  algorithm KAMER combined with First Fit as presented in
  \cite{bazargan00fast}}
\end{center}
\end{table*}

The running time of our algorithm is not only good in theory, but also
quite practical (as constants are small) and easy to implement. Here
we show some results of our implementation.  See Table~\ref{bench} for
an overview.

We randomly generated two different kinds of benchmark instances.
All of the instances describe a scenario in which 100 modules have to
be placed on an initially empty chip of size $80\times120$. Each
module stays on the chip until at least a certain number of new
modules have been placed. Then it is removed from the chip. This
number is different for each module and is randomly drawn from the
interval $[4, 100]$. Each of the modules needs to communicate with the
border of the chip and with all the modules located on the chip. The
buswidth is drawn for the interval $[0,10]$.

The instances differ in module size and distribution of the sizes. We
have generated instances where all modules have roughly the same size
(5-10\%, 10-15\%, 15-20\%, and 20-25\% of the chip size). These
instances are called uniform since the sizes are distributed
uniformly. We also have created instances where module sizes vary from
5 to 25\% of the chip size. Here we consider three different kinds of
distributions -- uniform, increasing, and decreasing. In the
increasing and decreasing case the modules are sorted by size.

Given these instances we benchmarked a g++ 3.2 compiled c++
implementation of our algorithm against the algorithms described in
\cite{ABT2004} and \cite{bazargan00fast}.  Shown in the first set of
columns in the table is a comparison of average running times for 100
modules for each instance in milliseconds on a 2.53GHz Intel Pentium 4
PC running under the linux operating system.  Remarkably, our
algorithm on average has the fastest running time, even though it
computes much better solutions. This illustrates the superiority of a
plane-sweep apporach.  Clearly, the difference in running times will
increase for even larger instances.

The second set of columns compares the average routing cost per
module. Routing costs are measured according to the weighted Manhattan
distance, which reflects the fact that routing on the chip is done in
an axis-parallel manner.  Note that in \cite{ABT2004}, placement is
done according to a weighted Euclidean distance, and optimization is
only done heuristically. As a consequence, the objective values are
markedly higher.  \cite{bazargan00fast} does not take routing cost
into account and places by some bin-packing like heuristic trying to
minimize rejection rate. This may result in modules being placed all
over the chip, regardless of communication cost. As a result,
communication cost is one order of magnitude higher than for our
method.

As a matter of fairness, we give a third set of columns, comparing the
average number of modules that had to be rejected due to lack of space
on the chip, which is one of the objectives in \cite{bazargan00fast}.
Note that this rejection rate does not play any role during the course
of our algorithm, nor is it considered in \cite{ABT2004}.  It is
striking that nevertheless, the total number of rejected modules for
our algorithm is precisely the same as for \cite{bazargan00fast}.
Again, our results dominate the ones for \cite{ABT2004} by a clear
margin.

In summary, our algorithm is faster except for case uniform15-20\%, better and more robust against
rejection than the method described in \cite{ABT2004}.  It is also
faster, much better and as robust against rejection as the approach
described in \cite{bazargan00fast}.


%% file: Conclusion.tex
\section{Conclusion}
\label{sec:Conclusion}

We have shown how to deal with two crucial issues supporting partial
reconfigurability in reconfigurable computing. This raises hope of
achieving further progress for even more complicated scenarios. One
such aspect is to streamline our data structures and algorithms for
repeated insertion or removal of modules. Some computational work can
be saved, as sorting from scratch is no longer required. While this
makes it relatively straightforward to lower the resulting complexity
of dealing with $n$ changes in total time $O(n^2)$, it remains an open
challenge to decide whether a subquadratic complexity is possible, as
no appropriate techniques for establishing quadratic lower bounds are
known. However, it is conceivable that we may be facing a 3SUM-hard
problem, which is the next best thing to an explicit lower bound. See
\cite{bh-pctmh-01} for techniques used for showing this for other
geometric problems.

An even more interesting challenge arises by venturing
from ``routing-con\-scious'' placement to ``routing-optimal''
placement: When routing among existing modules, we
may have to consider them as obstacles for our paths. Thus, distances
are not straight-line Manhattan distances, but {\em geodesic}
Manhattan distances, i.e., given by shortest paths among obstacles.
We believe that this type of problem can still be dealt with efficiently
by using the techniques of Mitchell~\cite{m-aaspt-88}.
This has been done in \cite{fmw-cwkmp-00} for an application to a 
routing-optimal placement problem for a continuum of demand points.
For dynamic routing requests at runtime, principles
that have been investigated include
Dynamic Network on Chip (DyNoC) \cite{bmakt-dnacr-04}
and Honeycomb \cite{tb-darrt-04}.

As our algorithm considers placing one module at a time,
it is an interesting problem to consider the more complex task
of placing the full set of modules at once. This is considered
in \cite{bbd+-capas-04} for the scenario of all processors
being of the same size; even without existing modules
and uniform routing cost, this turns out to be a tough
problem, as noted in \cite{KarpMcWo75}. We hope to provide
results on this scenario for modules of differing
size and non-uniform routing cost in the near future.

Finally, it should be interesting to consider placement of modules
as an online problem, where only limited information
is available at each stage. Interesting scenarios
require an appropriate modeling of the objective function
considered, in particular for the tradeoff between
computing cost, routing cost, and the cost of rejecting modules.
